\documentclass[aps,twocolumn,nofootinbib]{revtex4}
\usepackage[latin1]{inputenc}
\usepackage{epsfig}
\newcommand{\beq}{\begin{equation}}
\newcommand{\eeq}{\end{equation}}
\newcommand{\la}{\langle}
\newcommand{\ra}{\rangle}

\begin{document}

\title{Stochastic thermodynamics of system with continuous space of states}

\author{Mário J. de Oliveira}
\affiliation{Universidade de São Paulo, Instituto de Física,
Rua do Matão, 1371, 05508-090 São Paulo, SP, Brazil}

\begin{abstract}

We analyze the stochastic thermodynamics of systems with continuous
space of states. The evolution equation, the rate of entropy production,
and other results are obtained by a continuous time limit
of a discrete time formulation. We point out the role of time reversal
and of the dissipation part of the probability current on the production
of entropy. We show that the rate of entropy production is a bilinear
form in the components of the dissipation probability current with
coefficients being the components of the precision matrix related to
the Gaussian noise. We have also analyzed a type of noise that makes
the energy function to be strictly constant along the stochastic
trajectory, being appropriate to describe an isolated system.
This type of noise leads to nonzero entropy production and thus to an
increase of entropy in the system. This result contrasts with
the invariance of the entropy predicted by the Liouville equation,
which also describes an isolated system.

\end{abstract}

\maketitle

\section{Introduction}

The microscopic theory of systems in thermodynamic equilibrium
as advanced by Gibbs is based on the following assumptions.
An energy function is defined on the 
phase space, which is the space of the positions and velocities
of the elementary constituents of the system.  
A probability distribution is assigned to the phase space that
depends on the positions and velocity {\it only} through the energy
function. The entropy is directly related to the probability
distribution and is a generalization of the Boltzmann entropy.
As a consequence of these assumptions, the entropy becomes a function
of the mean energy from which it is possible to define temperature
by the Clausius relation, and derive the laws of 
equilibrium thermodynamics.

The Gibbs probability distribution does not properly characterize 
the thermodynamic equilibrium in a dynamic sense but is
a necessary condition for equilibrium.
The appropriate dynamic characterization of thermodynamic equilibrium 
is provided by the stochastic thermodynamics
\cite{tome2010,esposito2012,seifert2012,broeck2013,tome2015}.
Within this approach, thermodynamic equilibrium occurs when
the probability of occurrence of any trajectory equals the probability
of occurrence of its time-reversal trajectory. This condition
is also known as microscopic reversibility or detailed balance 
condition and is translated as the absence of 
entropy production.
As a consequence, the net current of any type, such as
heat current, will be absent, a property that provides
meaning to thermodynamic equilibrium in a dynamic sense. 

The distinguishing feature of the stochastic approach to thermodynamics
is the microscopic definition of the rate of entropy production.
Based on the macroscopic
bilinear relation between entropy production and thermodynamic
forces and affinities, Schnakenberg \cite{schnakenberg1976}
proposed a microscopic expression for the entropy production
of systems described by a master equation. 
The time variation of the entropy of these systems was shown to have two
parts, one of them being the production of entropy,
given by the Schnakenberg expression,
and the other being the entropy flux $\Phi$
\cite{jiuli1984,mou1986,lebowitz1999,crochik2005}.
The essential feature of the entropy production is its
straight relationship with the irreversibility
processes as expressed by the time-reversal symmetry
\cite{maes2003}. The entropy production is also directly 
related to probability current so that in a nonequilibrium
steady state these two quantities are nonvanishing 
\cite{zia2006,zia2007,andrieux2007,gaveau2009}.
The role of fluctuation theorems has also been addressed 
within the stochastic thermodynamics 
\cite{schmiedl2007,harris2007,seifert2008}.
The entropy production was calculated for molecular
motors \cite{andrieux2006}, in chemical reaction networks 
\cite{schmiedl2007,tome2018}, to determine the efficiency
at maximum power \cite{esposito2009}, and in systems connected
to multiple reservoirs \cite{tome2015,esposito2012}.
It was also determined in
irreversible interacting particle system where this quantity
was shown to display a singular behavior at the transition point
\cite{crochik2005,tome2012,artich2014,barbosa2018}.

A formulation of stochastic thermodynamics for continuous
system has also been developed, in which case the stochastic
evolution equation is the Fokker-Planck equation. 
It is assumed, usually in an implicit form, that the time-reversal
trajectory is identified as the reverse trajectory, which is also the
case of systems described by a master equation examined above.
This approach is appropriate for overdamped continuous systems
\cite{luposchainsky2013}. For one particle, 
the expression for the rate of entropy production is
proportional to the square of the probability current.
However, the application to a system that reaches a
non-equilibrium steady state, an extension of this
expression is needed and in fact, it has been advanced
\cite{sekimoto1998,tome2006,maes2008,broeck2010}. 

For underdamped continuous systems, the reverse trajectory
is no longer identified with the time-reversal trajectory and
an adequate formulation should be employed \cite{luposchainsky2013}.
For a system described by a Fokker-Planck-Kramers equation,
which is the stochastic equation appropriate for particles with
inertia, it has been found that the rate of entropy
production is related to just one part of the 
probability current \cite{chetrite2008,tome2010,spinney2012}, 
called, for this reason, the dissipation probability current.

The present approach describes underdamped systems, that is,
system consisting of particles with inertia, 
with continuous space of states.  
We focus on the production of entropy, understood as related to
the probability of occurrence of a trajectory and its time reversal.
When these two probabilities are equal we meet the condition
for the thermodynamic equilibrium. Defining the 
production of entropy as the logarithm of the ratio of these two
probabilities, it vanishes in thermodynamic
equilibrium. 

We consider systems consisting 
of interacting particles evolving according to the laws
of classical mechanics. In addition to the deterministic forces,
the system is also subject to random forces so that
the representative point in the space of states describes
a continuous stochastic trajectory. The deterministic force
is a sum of a time-reversal force and a force that lacks this
property and is identified as the dissipative force. 
The evolution equation is a continuity equation
for the probability density whose current is split into two
parts. One of them is the ordinary current related to the
time-reversal force. The other is the dissipative probability
current related to the dissipative force and the noise.

The evolution equation in the continuous space of states and
other properties are obtained by starting from a discrete time
formulation and then taking the continuous time limit.
In this sense the present method is distinct from the
previous similar methods \cite{spinney2012,luposchainsky2013}.
Our main result is the expression for the rate of entropy 
production obtained from a discrete time expression of the
production of entropy. The continuous time limit gives for
the rate of entropy production a bilinear form in the components
of the dissipative probability current which is positive definite. 
The vanishing of the dissipative probability current leads to
no entropy production characterizing the thermodynamic equilibrium.
 
We analyze in detail two types of noises. One of them is the
usual noise that describes the contact of a system with a heat
reservoir. The other type makes the energy function to be strictly
constant along a stochastic trajectory in phase space and thus
describes an isolated system. There is no flux of entropy and
the time variation of the entropy is entirely due to the generation
of entropy inside the system. This result is distinct from that given
by the Liouville equation which predicts an invariance of the entropy
in time and no production of entropy, although this equation
describes an isolated system.

It is convenient to regard the systems out of equilibrium as belonging 
in one of two classes. One of them includes the systems that are out of
equilibrium because they have not yet relaxed to the equilibrium state.
The other class includes those systems that are permanently out of
equilibrium even when they have already relaxed to the stationary state.
In this last case, entropy are permanently being produced by the system,
a feature that characterizes an out of equilibrium state.

Differently from the energy, which is a conserved quantity, the entropy
is not a conserved quantity but it cannot decrease, which is a brief
statement of the second law of thermodynamics.
Being a conserved quantity the increase of energy per unit time is given
by
\beq
\frac{dU}{dt} = \Phi_u,
\label{1}
\eeq 
where $\Phi_u$ is the rate at which energy is being introduced into the
system. The entropy increase per unit time on the other hand is given by
\beq
\frac{dS}{dt} =  \Pi - \Phi,
\label{2}
\eeq 
where $\Phi$ is the rate at which entropy is being delivered to outside
and $\Pi$ is the entropy production and obeys the inequality
$\Pi\ge0$, a brief statement of the second law of thermodynamics.

The approach we use here starts with the discrete expression
of the rate of the entropy production to reach the expression
for continuous systems by taking the continuous time limit.
Other approaches already consider the system to be continuous
in time and start from the expression for the entropy flux
defined as the heat flux divided by the temperature 
\cite{maes2008,maes2003a}, or start by identifying the production of
entropy as the relative entropy related to forward and
backward processes \cite{chetrite2008}.

\section{Evolution equation}

We consider a generic system whose state is defined as being the set
of variables $x_i$ understood as the components of a vector $x$ belonging
in a certain continuous space of states of a given dimension. As the system
evolves in time, the point representing the vector $x$ moves in the space of
states, tracing a trajectory. Supposing that the system is in a certain state
$x$ at time $t$, the question arises as to which trajectory the system will
follow starting at $x$. According to the stochastic assumption there is not
just one trajectory starting from $x$ but many possible trajectories,
each one occurring with a certain probability. 

To properly express the probability of occurrence of a
certain trajectory during a given interval of time 
it is necessary to specify not only the initial and final points
of the trajectory but also the intermediate points. These points are
understood as a time sequence of random variables and the probability
of the trajectory is a function of these variables. 
In addition, this probability could depend on previous states. 
However, according to the Markovian assumption adopted here,
the probability of a trajectory will not depend conditionally
on these other states. 
This assumption leads us to the conclusion that the probability of
the whole trajectory can be set up by specifying the probabilities 
of small sections of the trajectory. The probability of these elementary
trajectories dependent only on its initial and final points.

The probability of occurrence of an elementary trajectory that
starts within the elementary volume of the space of states $dx$
around the state $x$ and ends within $dx'$ around $x'$, after
a small interval of time $\tau$, is written as
\beq
P(x',x)dx'dx = K(x'|x)\rho(x)dx'dx,
\label{40a}
\eeq
where $\rho(x)dx$ is the probability of finding the system within
$dx$ around $x$ at a given time $t$ and $K(x'|x)dx'$ is the conditional
probability of finding the system within $dx'$ around $x'$ at time
$t+\tau$, given the occurrence of state $x$ at time $t$.

The main assumption of the present approach is that $x'$ is obtained from
$x$ by means of the following equation valid for small values of $\tau$ 
\beq
x_i' = x_i + F_i \tau + \xi_i \sqrt{\tau}, 
\label{33}
\eeq
where the forces $F_i(x)$ are given functions of $x$, and $\xi_i$
are random variables with a Gaussian distribution $G(\xi|x)d\xi$, understood
as a conditional probability, where $\xi$, the noise, denotes the vector with 
components $\xi_i$. The Gaussian distribution is such that the
random variables $\xi_i$ have zero means and covariances
$\la \xi_i\xi_j\ra=\Gamma_{ij}$.
The conditional probability distribution $K(x'|x)$ is obtained
from $G(\xi|x)$ by performing the transformation $\xi\to x'$
dictated by (\ref{33}). That is, the conditional probability $K(x'|x)$ of $x$
at time $t+\tau$ given $x$ at time $t$ is 
\beq
K(x'|x)dx' = G(\xi|x)d\xi,
\label{7}
\eeq
where the random variable $\xi$ is related to the random variable
$x'$ by (\ref{33}).

\begin{figure*}
\epsfig{file=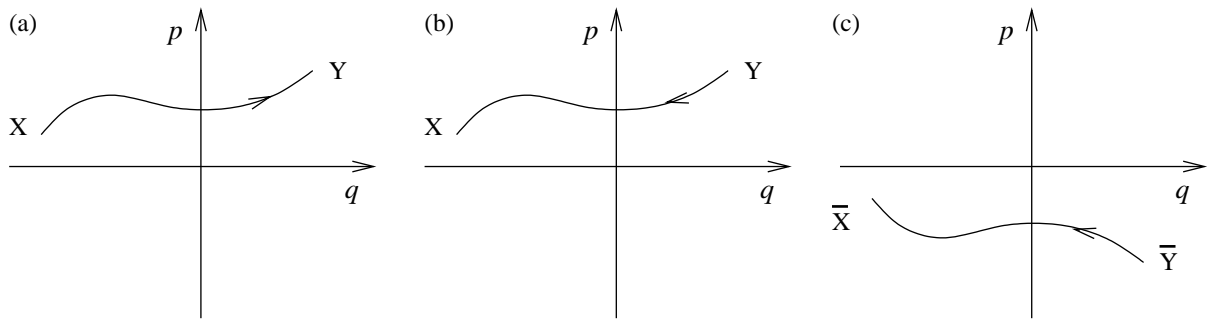,width=16cm}
\caption{Trajectories in the space of states, where $q$ is an even
variable and $p$ is an odd variable.
(a) The original trajectory, starting at X and
ending at Y. (b) The backward trajectory (but not time-reversed), that
starts at Y and ends at X.
(c) The time-reversal trajectory, that starts at $\bar{\mbox{Y}}$
and ends at $\bar{\mbox{X}}$. The points $\bar{\mbox{X}}$ and
$\bar{\mbox{Y}}$ are the time-reversal mappings of the 
points X and Y, respectively.}
\label{traj}
\end{figure*}

To find the continuous time equation, we start by denoting by $\rho'(x')$
the probability distributions at time $t+\tau$, and by $\rho(x)$ the
probability distribution at $t$. They are related to the conditional
probability $K$ through the equation
\beq
\rho'(x') = \int K(x'|x)\rho(x) dx,
\label{22}
\eeq
understood as the evolution equation for the probability
distribution in a discretized form. To find the evolution equation
in the continuous form, one should take the limit $\tau\to0$,
which is carried out as follows. We start by multiplying
both sides of equation (\ref{22}) by an arbitrary state function
${\cal F}(x')$ and integrate in $x'$,
\beq
\la {\cal F}\ra' = \int {\cal F}(x') K(x'|x)\rho(x) dx dx',
\eeq
where the average on the left-hand side is over the distribution $\rho'(x')$.
Changing the integration from $x'$ to $\xi$ the result is
\beq
\la {\cal F}\ra' = \int {\cal F}(x')G(\xi|x) \rho(x) dx d\xi,
\label{16}
\eeq
where here $x'$ is given by (\ref{33}).

Next we need the expansion of ${\cal F}(x')$ up to linear terms in $\tau$.
The expansion is obtained in two stages. First we expand this function
up to second powers of $\Delta x_i=x_i'-x_i$,
\beq
\Delta {\cal F} = \sum_i \frac{\partial{\cal F}}{\partial x_i}
\Delta x_i + \frac12 \sum_{ij} \frac{\partial^2{\cal F}}{\partial x_i x_j}
\Delta x_i \Delta x_j,
\eeq
where $\Delta {\cal F}={\cal F}(x')-{\cal F}(x)$.
Replacing the expressions (\ref{33}) into this equation
we reach the desired expansion
\beq
\Delta{\cal F} = \sum_i \frac{\partial{\cal F}}{\partial x_i}
(F_i\tau + \xi_i \sqrt{\tau})
+\frac12 \sum_i \frac{\partial^2{\cal F}}{\partial x_i x_j} \xi_i \xi_j \tau,
\label{26}
\eeq
valid up to terms of order $\tau$.

The expression (\ref{26}) is replaced in equation (\ref{16}) and
the integration in $\xi$ is carried out. Taking into account that
the average of $\xi_i$ vanishes, and that the average of
$\xi_i\xi_j$ is $\Gamma_{ij}$, the term proportional do $\sqrt{\tau}$
disappears and the whole right-hand side of the equation (\ref{16})
turns out to be proportional to $\tau$. After this procedure, we divide
both sides of the equation by $\tau$ to reach the result
\beq
\frac{d}{dt}\la {\cal F}\ra = \la {\cal K}^\dagger {\cal F}\ra,
\label{10}
\eeq
where we are considering
that $\Delta\la{\cal F}\ra/\tau \to \partial\la{\cal F}\ra/\partial t$
when $\tau\to0$, and ${\cal K}^\dagger$ is the differential operator
given by
\beq
{\cal K}^\dagger {\cal F} = \sum_i F_i \frac{\partial{\cal F}}{\partial x_i}
+ \frac12 \sum_{ij}\Gamma_{ij}\frac{\partial^2{\cal F}}{\partial x_i x_j},
\label{10a}
\eeq
and is the adjoint of the differential operator ${\cal K}$, defined by
\beq
{\cal K}\rho = -\sum_i \frac{\partial}{\partial x_i}(F_i \rho)
+ \frac12 \sum_{ij}\frac{\partial^2}{\partial x_i x_j}(\Gamma_{ij}\rho).
\label{21}
\eeq

Writing equation (\ref{10}) in the form
\beq
\int {\cal F}\frac{\partial\rho}{\partial t}dx
=  \int {\cal F}({\cal K} \rho) dx,
\eeq
obtained by appropriate integrations by parts and by taking into 
account that $\rho$ vanishes rapidly in the limits of integration,
we conclude that
\beq
\frac{\partial \rho}{\partial t} = {\cal K}\rho,
\eeq
or in an explicit form,
\beq
\frac{\partial\rho}{\partial t}
= -\sum_i \frac{\partial}{\partial x_i}( F_i \rho)
+ \frac12 \sum_{ij}\frac{\partial^2}{\partial x_i \partial x_j}
(\Gamma_{ij}\rho),
\label{14}
\eeq
which is the desired equation that gives the time evolution of the
probability distribution $\rho(x,t)$ in a continuous form, and is a
Fokker-Planck equation \cite{kampen1981,gardiner1983,risken1984,tome2015L}.

\section{Production of entropy}

\subsection{Time reversal and entropy production}

Irreversible processes are characterized by the lack of
time-reversal invariance which means that the probability
of the occurrence of a certain process is different from the
probability of its time reversal. In accordance with thermodynamics,
a measure of irreversibility is how much entropy is being
generated. Thus the production of entropy is directly related 
to the lack of time reversibility.

Given a trajectory in the space of states, the time-reversal trajectory
may not be, generally speaking, its reverse, as illustrated in
figure \ref{traj}. If a trajectory starts at the point 
$x$ and ends at $x'$, the reverse starts at $x'$ and ends at $x$,
and may not coincide with the time-reversal trajectory which is
understood as follows. Let $x\to \bar{x}$ be a mapping that associates
to each state $x$ a time-reversal state $\bar{x}$.
If $x$ and $x'$ are the initial and final states of a trajectory then
the initial and final states of the time-reversal trajectory are,
respectively, $\bar{x}'$ and $\bar{x}$. That is, the final state of
the original trajectory maps onto the initial state of the time-reversal
trajectory and vice-versa.

The type of time-reversal mapping that we consider is 
such that $x_i$ either changes its sign or keep its sign in 
the transformation $x\to\bar{x}$. It is thus convenient to 
classify the variables $x_i$ into two categories.
If $x_i$ keeps its sign it belongs in the first category or is
of the even type. If $x_i$ changes sign, it belongs in the
second category or is of the odd type. It is worth mentioning that
if $\bar{x}_i \bar{x}_j = x_i x_j$, then $x_i$ and $x_j$ belong 
in the same category, otherwise they belong in distinct categories.

The time reversal of a vector state function such as the force $F$
is defined in terms of its components. The time reversal of
$F_i$ is denoted $\bar{F}_i$ and equals
$F_i$ or $-F_i$ according to whether $x_i$ is of the even or odd type,
respectively. 

In general, the probability of occurrence of a certain trajectory
$x\to x'$, during a small interval of time $\tau$, which is
\beq
P(x',x) = K(x'|x)\rho(x),
\eeq
is different from the probability of occurrence of the time-reversal
trajectory $\bar{x}'\to\bar{x}$, which is 
\beq
P(\bar{x},\bar{x}') = K(\bar{x}|\bar{x}')\rho(\bar{x}').
\label{40b}
\eeq
A very special situation occurs when the probability of a trajectory
and its time reversal is equal. Thermodynamic equilibrium corresponds
to the case when this equality occurs for all trajectories.
A measure of the departure from equilibrium may be given by
the logarithm of the ratio of these two probabilities,
\beq
\ln\frac{P(x',x)}{P(\bar{x},\bar{x}')},
\eeq
a quantity that vanishes when the two probabilities are equal.
We must integrate over all possible trajectories occurring 
during the interval of time $\tau$, leading us to the 
following expression for the production of entropy during 
the interval of time $\tau$,
\beq
\int P(x',x)\ln\frac{P(x',x)}{P(\bar{x},\bar{x}')} dxdx'.
\label{27}
\eeq

The rate of production of entropy $\Pi$ is defined by dividing 
(\ref{27}) by $\tau$ and by multiplying by the Boltzmann constant $k$, 
\beq
\Pi = \frac{k}{\tau}\int P(x',x)
\ln\frac{P(x',x)}{P\bar{x},\bar{x}')}dx'dx,
\label{39}
\eeq
and it is understood that we should take the limit $\tau\to0$.
Writing this equation in the equivalent form
\beq
\Pi = \frac{k}{2\tau}\int \{P(x',x) - P(\bar{x},\bar{x}')\}
\ln\frac{P(x',x)}{P(\bar{x},\bar{x}')}dx' dx,
\eeq
it becomes clear that $\Pi\ge0$ because the integrand is never negative.
In terms of the conditional probability, the rate of entropy production
reads
\[
\Pi = \frac{k}{2\tau}\int \{ K(x'|x)\rho(x) - K(\bar{x}|\bar{x}')
\rho(\bar{x}')\} \times
\]
\beq
\times\ln\frac{K(x'|x)\rho(x)}{K(\bar{x}|\bar{x}')\rho(\bar{x}')}dx' dx.
\label{31}
\eeq
For the discrete space of states, the integral is replaced by a summation,
in which case this expression becomes the expression proposed by
Schnakenberg for the production of entropy related to a master
equation \cite{schnakenberg1976}.

The expresion (\ref{31}) is not the entropy $S$ of the system, which
is defined by
\beq
S = -k \int \rho(x) \ln \rho(x) dx,
\eeq
and, in general, it is not either the variation of the entropy
with time $dS/dt$, which is
\beq
\frac{dS}{dt} = \frac{k}{2\tau}\int
\{ K(x'|x)\rho(x) - K(\bar{x}|\bar{x}')\rho(\bar{x}')\}
\ln\frac{\rho(x)}{\rho(\bar{x}')}dx' dx,
\label{45}
\eeq
where we assumed that $\rho(\bar{x})=\rho(x)$.
The difference $\Phi=\Pi-dS/dt$ is given by
\beq
\Phi = \frac{k}{2\tau}
\int \{ K(x'|x)\rho(x) - K(\bar{x}|\bar{x}')\rho(\bar{x}')\}
\ln\frac{K(x'|x)}{K(\bar{x}|\bar{x}')}dx' dx,
\label{46}
\eeq
and is interpreted as the flux of entropy per unit time {\it from}
the system {\it to} the outside.

\subsection{Rate of entropy production}

Next we wish to determine the rate of entropy production
in the limit $\tau\to0$.  
We recall that the conditional probability $K(x'|x)$ is related to the 
noise probability distribution by relation (\ref{7}),
where $G(\xi|x)$ is the probability distribution of the noise $\xi$,
related to $x'$ by
\beq
\xi_i = \frac1{\sqrt{\tau}}(x_i'-x_i - F_i(x)\tau), 
\label{36}
\eeq
where $F_i$ are functions of $x$.

We assume that the noises $\xi_i$ are distributed according to the
Gaussian distribution $G(\xi|x)$ in several variables, with zero means
and covariances $\la \xi_i\xi_j\ra=\Gamma_{ij}$ that may depend on $x$.
Given the covariances, the Gaussian distribution is uniquely determined
and is given by
\beq
G(\xi|x) = \frac1Z \exp\{- \frac12 \sum_{ij} \xi_i B_{ij}\xi_j \},
\label{43}
\eeq
where
\beq
Z = \int \exp\{- \frac12 \sum_{ij} \xi_i B_{ij}\xi_j \} d\xi,
\label{47}
\eeq
and $B$, the matrix with elements $B_{ij}$, is the
inverse of the covariance matrix $\Gamma$, and may depend on $x$.
As $G(\xi|x)$ describes a probability distribution, the eigenvalues
of the precision matrix $B$ and
of the covariant matrix $\Gamma$ are greater or equal to zero.

To determine the rate of entropy production,
we write (\ref{31}) in terms of the Gaussian distribution by the
use of (\ref{7}) and by employing the conditional probability
$K(\bar{x}|\bar{x}')$ related to the time-reversal trajectory,
\beq
K(\bar{x}|\bar{x}')d\bar{x} = G(\xi^*|\bar{x}')d\bar{\xi}',
\label{7b}
\eeq
where $\xi^*$ is given by 
\beq
\xi_i^* = \frac1{\sqrt{\tau}}(\bar{x}_i-\bar{x}_i' - F_i(\bar{x}')\tau).
\label{36a}
\eeq
Notice that the right-hand side of (\ref{36a}) is not the time reversal
of the right-hand side of (\ref{36}). For this reason, we are using
the notation $\xi_i^*$ and not $\bar{\xi}_i$.
In terms of the Gaussian distribution, the rate of entropy production reads
\[
\Pi = \frac{k}{2\tau}\int \{G(\xi|x)\rho(x)
- G(\xi^*|\bar{x}')\rho(\bar{x}') \} \times
\]
\beq
\times
\ln\frac{G(\xi|x)\rho(x)}{G(\xi^*|\bar{x}')\rho(\bar{x}')}d\xi dx,
\label{31a}
\eeq
which is obtained by a change of variables from $x'$ to $\xi$,
given by (\ref{36}) and we remark that $\xi_i^*$ is related to
both $x'$ and $x$ by (\ref{36a}) so that all terms in the integrand
involve only the variables $\xi$ and $x$.

Before we start the calculation, we assume two properties of the
covariances, the denial of which would lead to an artificial production
of entropy. The first property is
\beq
\Gamma_{ij}(\bar{x})=\Gamma_{ij}(x),
\label{01}
\eeq
and is valid also for $B_{ij}(x)$, $Z(x)$ and $\rho(x)$.
The second property is that $\Gamma_{ij}(x)$ vanishes whenever 
$x_i$ and $x_j$ belong in distinct categories, that is, if one is
even and the other is odd, and is also valid for $B_{ij}(x)$.
This property is conveniently written as
\beq
\Gamma_{ij}\bar{x}_i\bar{x}_j = \Gamma_{ij}x_ix_j.
\label{02}
\eeq

\subsection{Additive noise}

We consider here the case in where the covariant matrix
$\Gamma$ does not depend on $x$, and the same is valid for
the precision matrix $B$. We start by expanding the expression 
\beq
\ln\frac{G(\xi|x)\rho(x)}{G(\xi^*|\bar{x}')\rho(\bar{x}')},
\label{29}
\eeq
up to terms of order $\sqrt{\tau}$. Using the definition of the
Gaussian distribution, this expression may be written as
\beq
 \frac12 \sum_{ij} B_{ij}(\xi_i^* \xi_j^* - \xi_i\xi_j)
- \ln\frac{\rho(x')}{\rho(x)},
\label{85}
\eeq
where we used the property (\ref{01}) for $B_{ij}$ and $\rho$.

To determine the first term of (\ref{85}), we observe that 
in accordance with the property (\ref{02}), valid for $B_{ij}$,
\beq
B_{ij}\xi_i^*\xi_j^* = B_{ij}\bar{\xi_i^*}\bar{\xi_j^*},
\eeq
where $\bar{\xi_i^*}$ is the time reversal of $\xi_i^*$, 
\beq
\bar{\xi_i^*} = \frac1{\sqrt{\tau}}(x_i - x_i' - \bar{F}_i(\bar{x}')\tau),
\eeq
so that, up to terms of order $\sqrt{\tau}$,
\beq
\bar{\xi_i^*} = - \xi_i - 2D_i(x)\sqrt{\tau},
\eeq
where
\beq
D_i(x) = \frac12[\bar{F}_i(\bar{x}) + F_i(x)].
\eeq
The first term of (\ref{85}), up to terms of order $\sqrt{\tau}$, becomes
\beq
\sum_{ij} B_{ij}[\xi_i D_j(x) + \xi_j D_i(x)]\sqrt{\tau}.
\eeq

Considering that up to terms of order $\sqrt{\tau}$,
$x_i' = x_i + \xi_i \sqrt{\tau}$, and using the property (\ref{01})
for $\rho$, the second term of (\ref{85}) becomes 
\beq
- \sum_k \frac{\partial\ln\rho}{\partial x_k}\xi_k \sqrt{\tau}.
\eeq

Collecting these results, we may write
\beq
\ln\frac{G(\xi|x)\rho(x)}{G(\xi^*|\bar{x}')\rho(\bar{x}')}
= {\cal A}(\xi,x) \sqrt{\tau},
\label{73}
\eeq
where
\beq
{\cal A} =  \sum_i {\cal A}_i \xi_i,
\label{58a}
\eeq
and
\beq
{\cal A}_i =  2\sum_j B_{ij} D_j - \frac{\partial\ln\rho}{\partial x_i}.
\label{71}
\eeq

In a similar fashion we find
\beq
G(\xi|x)\rho(x) - G(\xi^*|\bar{x}')\rho(\bar{x}') = G(\xi|x)\rho(x)
{\cal A}(\xi,x) \sqrt{\tau},
\eeq
and the rate of entropy production becomes
\beq
\Pi = \frac{k}{2}\int G(\xi|x)\rho(x)[{\cal A}(\xi,x)]^2 d\xi dx.
\label{31b}
\eeq

Replacing the result (\ref{58a}) for ${\cal A}$ in the expression
(\ref{31b}), performing the integral in $\xi$, and bearing in mind
that $\la \xi_i\xi_j\ra=\Gamma_{ij}$, we reach the following desired
result for the rate of entropy production,
\beq
\Pi = \frac{k}2 \sum_{ij}
\int {\cal A}_i \,\Gamma_{ij}{\cal A}_j \,\rho \, dx, 
\label{69}
\eeq
which is clearly nonnegative because the eigenvalues of $\Gamma_{ij}$ 
are nonnegative.

Comparing equations (\ref{45}) and (\ref{31}), we observe that 
they differ from the last factor in the integrand of both equations.
An expression for $dS/dt$ can thus be obtained by using the same
reasoning that led us from (\ref{31}) to (\ref{69}).
The result is
\beq
\frac{dS}{dt} = - \frac{k}2 \sum_{ij} \int {\cal A}_i \,
\Gamma_{ij}\frac{\partial\rho}{\partial x_j} \, dx.
\label{70}
\eeq
To find an expression for the flux of entropy $\Phi$, we recall that
$\Phi=\Pi-dS/dt$. Subtracting the expressions (\ref{69}) and (\ref{70}),
we get 
\beq
\Phi = k \sum_i \int {\cal A}_i D_i \,\rho \, dx,
\label{70a}
\eeq
where we used the relation $B\Gamma=I$.

\section{Probability current}

\subsection{Dissipation probability current}

The evolution equation (\ref{14}) can be written in the following form
\beq
\frac{\partial\rho}{\partial t}
= -\sum_i \frac{\partial J_i^c}{\partial x_i},
\label{28}
\eeq
where
\beq
J_i^c = F_i \rho - \frac12 \sum_j\frac{\partial}{\partial x_j}
(\Gamma_{ij}\rho). 
\eeq
In this form, the evolution equation is a continuity equation and
$J_i^c$ is the probability current. Next, we wish to split the probability
currents into two parts, one of them being invariant under time reversal.
To this end, we consider first the splitting of the force $F_i$.

Any force $F_i(x)$ can always be split into two parts, one of them being
\beq
D_i(x) = \frac12[F_i(x) + \bar{F}(\bar{x})],
\eeq
and the other being
\beq
F_i^r(x) = \frac12[F_i(x) - \bar{F}(\bar{x})].
\eeq
That is, 
\beq
F_i(x) = F_i^r(x) + D_i(x).
\eeq
The first part $F_i^r$ is invariant under time reversal,
holding the time-reversal property
\beq
\bar{F}_i^r(x)=-F_i^r(\bar{x}).
\label{67a}
\eeq
In an explicit form, if $F_i^r$ is an odd type of force, which is
identified as an ordinary force, the time-reversal property reads
$F_i^r(x)=F_i^r(\bar{x})$. If $F_i$ is an even type of force, the
time-reversal property reads $F_i^r(x)=-F_i^r(\bar{x})$.
From (\ref{67a}), it follows that
$A_i=\partial F_i^r/\partial x_i$ holds the property
\beq
A_i(\bar{x}) = - A_i(x).
\label{89}
\eeq

The second part $D_i$ is the dissipative part, which holds the property
\beq
\bar{D}_i(x)=D_i(\bar{x}).
\label{67b}
\eeq
If $D_i$ is an odd type of force,
this property reads, $D_i(x)=-D_i(\bar{x})$, 
and $D_i$ is identified with a dissipative
force, an example of which is the ordinary dissipation proportional
to the velocity.  If $D_i$ is an even type of
force, this property reads $D_i(x)=D_i(\bar{x})$.
Only the second part, $D_i$, that lacks the time-reversal property,
contributes to the production of entropy as can be observed
by looking at equations (\ref{71}) and (\ref{31b}).

In an analogous manner, the probability current is split into two parts
\beq
J_i^c = J_i^r + J_i,
\eeq
where the first part is the reversible probability current,
\beq
J_i^r = F_i^r \rho
\eeq
which is invariant under time reversal, holding the property (\ref{67a})
because $\rho(\bar{x})=\rho(x)$, and the second part is the
irreversible probability current, 
\beq
J_i = D_i \rho - \frac12 \sum_j
\frac{\partial \rho\Gamma_{ij}}{\partial x_j},
\label{74}
\eeq
which holds the property (\ref{67b}) because
$\Gamma_{ij}(\bar{x})=\Gamma_{ij}(x)$.

\subsection{Time variation of the entropy}

The variation of the entropy 
\beq
S = - k \int \rho \ln\rho dx,
\eeq
with time is 
\beq
\frac{dS}{dt} = - k \int \frac{\partial\rho}{\partial t} \ln\rho dx.
\label{75}
\eeq
Using the evolution equation in the form (\ref{28}),
it can be written as
\beq
\frac{dS}{dt}
= k \sum_i \int \frac{\partial J_i^c}{\partial x_i} \ln\rho \,dx,
\label{70b}
\eeq
Replacing $J_i^c$ by $J_i^r+J_i$, the right-hand side will
be a sum of two terms, one of which involves the integral
\beq
\sum_i\int \frac{\partial J_i^r}{\partial x_i} \ln\rho \,dx
= \sum_i\int \frac{\partial F_i^r}{\partial x_i} \rho\,dx,
\eeq
where the equality was obtained by two integrations by parts. 
But this expression vanishes in view of the property (\ref{89})
and we are left only with the second part, 
\beq
\frac{dS}{dt} = 
- k \sum_i \int J_i \frac{\partial \ln\rho}{\partial x_i} dx,
\label{75a}
\eeq
where an integration by parts has been performed.

If we define $F_i^{\rm ir}=J_i/\rho$, we may write, after an
integration by parts,
\beq
\frac{dS}{dt} = 
k \sum_i \int \frac{\partial F_i^{\rm ir}}{\partial x_i} \rho \,dx,
\eeq
In this form we see that the time variation of the entropy
is related to the change in the volume of phase space, 
measured by the divergence of $F^{\rm ir}$.

\subsection{Rate of entropy production}

The comparison of equations (\ref{75a}) and (\ref{70}) indicates
that $J_i$ is related to ${\cal A}_i$ by
\beq
J_i = \frac\rho2\sum_j {\cal A}_j \,\Gamma_{ij}.
\eeq
Inverting this relation, we find
\beq
{\cal A}_i = \frac2\rho\sum_j B_{ij}J_j,
\eeq
where we used $B\Gamma=I$, 
which leads us to the following expression
\beq
{\cal A}_i = 2 \sum_j B_{ij} D_j - \frac{\partial \ln\rho}{\partial x_i}
- \sum_{jk} B_{ij}\frac{\partial\Gamma_{jk}}{\partial x_k},
\label{77}
\eeq
obtained by using (\ref{74}), where again we used $B\Gamma=I$.

We have seen above that the rate of entropy production is given by
expression (\ref{69}), which was demonstrated to be the rate
of entropy for the case in which $\Gamma_{ij}$ does not depend on $x$,
in which case the expression (\ref{77}) for ${\cal A}_i$ does not have
the last term on the right-hand. Although we did not show that the
expression (\ref{69}) is also valid for the
case in which $\Gamma_{ij}$ depends on $x$, we assume that it
expresses the rate of entropy production in this case, with ${\cal A}_i$
given by (\ref{77}).

Using the relation between ${\cal A}_i$ and
$J_i$, the rate of entropy production can be written in terms
of the dissipation probability current as
\beq
\Pi = k\sum_i \int J_i {\cal A}_i dx,
\label{69d}
\eeq
or as 
\beq
\Pi = 2 k\sum_{ij} \int \frac1\rho J_i B_{ij}J_j dx. 
\label{69c}
\eeq
This expression is clearly nonnegative because the
eigenvalues of $B$ are nonnegative and we notice that
it is related only to the dissipation
part of the probability current. When $B$ is diagonal,
this formula was considered by Tomé and de Oliveira
\cite{tome2010} and derived by Spinney and Ford \cite{spinney2012}
by a method which has similarities with the present approach.
The expression (\ref{69c}) was derived by Chetrite and
Gaw\c{e}dzki \cite{chetrite2008} by identifiying the production
of entropy as a relative entropy related to forward and
backward processes.

The flux of entropy $\Phi$ is obtained by recalling that $\Phi=\Pi-dS/dt$.
Subtracting the expressions (\ref{69c}) and (\ref{70b}), we get 
\beq
\Phi = 2 k \sum_{ij} \int J_i B_{ij} L_j dx,
\label{88}
\eeq
where
\beq
L_j = D_j - \frac12 \sum_k \frac{\partial \Gamma_{jk}}{\partial x_k},
\label{88a}
\eeq 
which can also be written as
\beq
\Phi = k\sum_i \int {\cal A}_i L_i \rho \,dx.
\eeq

\section{Energy, heat and work}

From now on, we wish to describe a system that may be acted by internal
as well as by external forces. The internal forces are considered to be
conservative forces in the sense that they are derived from an energy
function $E(x)$ associated to the system. 
Let $x_i$ and $x_j$ be a pair of even and odd variables, respectively.
Then the even conservative force $F_i^c$ and
the odd conservative force $F_j^c$ are obtained from
the energy function $E(x)$ by 
\beq
F_i^c = \frac{\partial E}{\partial x_j}, \qquad\qquad
F_j^c =-\frac{\partial E}{\partial x_i}.
\label{86}
\eeq
The energy function holds the time-reversal property,
$E(\bar{x})=E(x)$, guaranteeing the time-reversal property
(\ref{67a}) of the conservative forces.

In addition to the internal forces $F_i^c$, the system,
if it is not isolated, may be acted by external forces $F_i^e$
which are also considered to be time reversal.
The force $F_i^r$ becomes a sum of these two forces
\beq
F_i^r = F_i^c + F_i^e,
\eeq
and the evolution equation (\ref{14}) becomes
\beq
\frac{\partial\rho}{\partial t} =
- \sum_i \frac{\partial F_i^c\rho}{\partial x_i}
- \sum_i \frac{\partial F_i^e\rho}{\partial x_i} 
- \sum_i\frac{\partial J_i}{\partial x_i}.
\label{14c}
\eeq

From the property (\ref{86}), it follows at once the following result
\beq
\sum_i \frac{\partial F_i^c}{\partial x_i} = 0.
\eeq
Using this property, we find
\beq
\sum_i \frac{\partial F_i^c\rho}{\partial x_i}
= \sum_i F_i^c\frac{\partial \rho}{\partial x_i},
\eeq
which can be written as
\beq
- \sum_i F_i^c\frac{\partial \rho}{\partial x_i} 
= \sum_{(ij)} \left(\frac{\partial E}{\partial x_i}
\frac{\partial\rho}{\partial x_j} - \frac{\partial E}{\partial x_j}
\frac{\partial\rho}{\partial x_i}\right) =\{E,\rho\},
\eeq
where the summation extends over all pairs $(i,j)$ such that $x_i$
and $x_j$ consist of a pair of conjugate variables such that the
$x_i$ is even and $x_j$ is odd, and this summation is recognized
as the Poisson brackets between $E$ and $\rho$.

The evolution equation (\ref{14c}) then becomes
\beq
\frac{\partial\rho}{\partial t}
= \{E,\rho\} -\sum_i \frac{\partial F_i^e \rho}{\partial x_i}
- \sum_i\frac{\partial J_i}{\partial x_i}.
\label{14b}
\eeq
The time evolution of the average of the energy $\la E(x)\ra$,
understood as the thermodynamic internal energy $U$ of the system,
is obtained by multiplying (\ref{14b}) by $E(x)$ and integrating in $x$.
The result is 
\beq
\frac{dU}{dt} = \sum_i \int J_i \frac{\partial E}{\partial x_i} dx
+ \sum_i \int F_i^e \rho\frac{\partial E}{\partial x_i} dx,
\label{91a}
\eeq
obtained after appropriate integrations by parts. The first summation on
the right hand-side is identified as the total heat flux introduced into
the system,
\beq
\Phi_q = \sum_i \int J_i \frac{\partial E}{\partial x_i} dx,
\label{91b}
\eeq
and the second as minus the work performed by the system per unit time,
or power generated by the system,
\beq
\Phi_w = - \sum_i \int F_i^e\rho \frac{\partial E}{\partial x_i} dx.
\eeq
The equation (\ref{91a}) acquires the form
\beq
\frac{dU}{dt} = \Phi_q - \Phi_w,
\eeq
which is understood as the global conservation of energy,
and $\Phi_u$ in equation (\ref{1}) is $\Phi_u=\Phi_q-\Phi_w$.

\section{A special type of noise}

The noise, which is represented by the covariances matrix $\Gamma$ 
is not yet fully specified. Some of their essential properties
have already been presented in equations (\ref{01}) and (\ref{02}), 
and are: $\Gamma_{ij}(\bar{x})=\Gamma_{ij}(x)$;
and $\Gamma_{ij}(x)$ vanishes whenever $x_i$ and $x_j$ consists of a
pair of even and odd types. There are many choices of noise depending on
the physical situation one wants to describe. Here we take a look at the
type of noise that leaves a certain quantity $E(x)$ invariant along the
trajectory determined by this noise. The quantity $E$ is strictly
constant in every possible stochastic trajectory, and not only
on the average. If two states $x$ and $x'$ are related by
\beq
x_i' = x_i + F_i \tau + \xi_i \sqrt{\tau},
\eeq
then the expansion of $E(x')-E(x)$ up to terms of order $\tau$ is
\beq
E(x')-E(x) =
\sum_i\frac{\partial E}{\partial x_i} (F_i \tau + \xi_i \sqrt{\tau})
+ \frac12 \sum_{ij}
\frac{\partial^2 E}{\partial x_i\partial x_j} \Gamma_{ij} \tau,
\eeq
where as before $\Gamma_{ij}$ denotes the covariance of the
random variables $\xi_i$.

If $E(x')=E(x)$ along the trajectory then the following
constraint should be obeyed
\beq
\sum_j \xi_j f_j = 0,
\label{57}
\eeq
where
\beq
f_j=\frac{\partial E}{\partial x_j},
\label{57a}
\eeq
and
\beq
\sum_i f_i F_i
+ \frac12 \sum_{ij} \Gamma_{ij} \frac{\partial f_i}{\partial x_j} = 0.
\label{58}
\eeq
The first condition means that the random variables are not independent
variables but are connected by (\ref{57}). Multiplying (\ref{57}) by
$\xi_i$ and taking the average over the random variable $\xi$, we find
\beq
\sum_j \Gamma_{ij} f_j = 0,
\label{62}
\eeq
which relates the covariances and $f_i$. 
Owing to the relation (\ref{62}), the condition (\ref{58}) 
is equivalently expressed by
\beq
F_i = \frac12 \sum_j \frac{\partial\Gamma_{ij}}{\partial x_j}.
\label{63}
\eeq
If a certain quantity remains constant along a stochastic
trajectory, the random variables $\xi_i$ should be connected
by (\ref{57}), and $F_i$ should be related to the covariances by
(\ref{63}).

Replacing the condition (\ref{63}) in equation (\ref{88a}), we
see that the quantity $L_i$ vanishes and so does the flux of entropy,
given by (\ref{88}). In other terms, the flux of entropy vanishes for
the conservative noise that we are considering here and one concludes
from this property that the variation of the entropy of the system
$dS/dt$ equals the rate of the entropy production $\Pi$.

A noise that meet the condition (\ref{57}) is set up as follows.
For $i\neq j$, let $\xi_{ij}$ be random variables with zero means,
each one with variance $\lambda_{ij}=\lambda_{ji}\geq0$,
that is, $\la \xi_{ij}^2\ra=\lambda_{ij}$. These are independent
random variables, except $\xi_{ij}$ and $\xi_{ji}$ which are related by
\beq
\xi_{ji}=-\xi_{ij}.
\label{83}
\eeq
The random variable $\xi_i$ is defined in terms
of these new random variables by
\beq
\xi_i = \sum_{j(\neq i)} \xi_{ij} f_j.
\label{84}
\eeq
Using property (\ref{83}), the condition (\ref{58}) follows immediately.
We recall that $f_i=\partial E/\partial x_i$ and may depend on $x$,
where $E(x)$ is the conserved quantity.

From (\ref{84}) we may determine the covariances
$\Gamma_{ij}=\la\xi_i\xi_j\ra$. Using the property (\ref{83}) we find
\beq
\Gamma_{ij} = -\lambda_{ij}f_if_j,  
\eeq
for $i\neq j$, and
\beq
\Gamma_{ii} = \sum_{j(\neq i)} \lambda_{ij} f_j^2.
\eeq
From these results, we see that (\ref{62}) is verified.

\section{Thermodynamic equilibrium}

\subsection{Noise-dissipation relation}

From now on we consider only the situations such that the external
forces are not present, in which case the evolution equation is
\beq
\frac{\partial\rho}{\partial t}
= \{E,\rho\} - \sum_i\frac{\partial J_i}{\partial x_i}.
\label{24a}
\eeq
It remains to choose which type of noise to use.
The choice of noise, represented by the covariances $\Gamma_{ij}$,
and of the dissipative forces $D_i$ is guided by the type
of situation one wants to describe. If we wish to describe 
an equilibrium situation, the noise represented by the covariances
$\Gamma_{ij}$ and the dissipation represented by $D_i$ 
cannot be arbitrary but must hold a relationship between them,
a noise-dissipation relation.

For long times, the density
$\rho$ will reach a stationary density $\rho_s$, which makes the
right-hand side of equation (\ref{24a}) to vanish. If $J_i(\rho_s)$
is nonzero for some $i$, then $\Pi$ is nonzero and the stationary
state will be a state in which entropy is continuously been produced,
and this is not an equilibrium state. The thermodynamic equilibrium is
characterized by the vanishing of the entropy production which implies
that $J_i$ should vanish for all $i$. Denoting by $\rho_e$ the equilibrium
probability distribution then the condition for thermodynamic equilibrium is
\beq
J_i(\rho_e)=0,
\label{97a}
\eeq
for all $i$. Recalling the definition of $J_i$, given by (\ref{74}),
this condition is equivalent to
\beq
D_j - \frac12 \sum_k \frac{\partial\Gamma_{jk}}{\partial x_k}
= \frac12 \sum_{k} \Gamma_{jk} \frac{\partial \ln \rho_e}{\partial x_k},
\label{90}
\eeq
for all $i$.

Let us analyze the types of covariances $\Gamma_{ij}$ and the dissipative
force $D_i$ that may lead the system to the thermodynamic equilibrium.
As the quantity $J_i(\rho_e)$ vanishes for each $i$, the second summation
on the right-hand side of equation (\ref{24a}) disappears and the first
summation must vanish as well, that is, 
\beq
\{E,\rho_e\} = 0.
\label{97b}
\eeq
This equation is fulfilled if $\rho_e$ is a function of $E$, that is if
$\rho_e(x)=\rho(E(x))$ depends on $x$ through the energy function $E(x)$.
In other words, in the thermodynamic equilibrium, the probability density
is a function of the energy function, which is the main property of 
the equilibrium Gibbs distributions.
The general condition for thermodynamic equilibrium is reduced to the
condition represented by equation (\ref{90}) where $\rho_e$ is understood
as a function of the energy function $E(x)$. With this understanding,
the equation (\ref{90}) is the noise-dissipation relation.

The equations (\ref{97a}) and (\ref{97b}) are the two conditions
that gives the equilibrium probability distribution. The first
condition represents the detailed balance condition or microscopic
reversibility and the second is related to the conservation of
energy. These two conditions are the ones used implicitly by
Maxwell in his second derivation of the velocity distribution that
bears his name \cite{oliveira2019}.

\subsection{Canonical setting}

Let us consider two relevant cases. The first is the one
in which $\rho_e$ is proportional to $e^{-\beta E}$,
which corresponds to the Gibbs canonical distribution.
In this case equation (\ref{90}) reduces to 
\beq
D_j - \frac12 \sum_k \frac{\partial\Gamma_{jk}}{\partial x_k}
= -\frac{\beta}2 \sum_{k} \Gamma_{jk} \frac{\partial E}{\partial x_k},
\label{90a}
\eeq
which is the noise-dissipation relation for the present case.

Using relation (\ref{90}),
the flux of entropy (\ref{88}) reduces to the following simple form 
\beq
\Phi = k \sum_i \int J_i \frac{\partial \ln \rho_e}{\partial x_i} dx
= -\frac1T \sum_i \int J_i \frac{\partial E}{\partial x_i} dx.
\label{88b}
\eeq
The comparison of the expressions (\ref{88b}) and (\ref{91b}),
leads us to the relation 
\beq
\Phi=-\frac{\Phi_q}T,
\eeq
which connects the flux of entropy and the heat flux. Since
$dU/dt=\Phi_q$ and $dS/dt=\Pi-\Phi$, we reach the relation
\beq
\frac{dS}{dt} = \Pi + \frac1T\frac{dU}{dt}.
\label{87}
\eeq
Near equilibrium, the rate of entropy production vanishes and
we are left with the relation $dU=TdS$, which confirms
that the noises and dissipation satisfying the noise-dissipation 
relation (\ref{90a}) describe a system in contact
with a reservoir at a temperature $T$.

If the temperature is kept constant, then the variation with
time of the free energy $F=U-TS$ is related to the entropy
production by $dF/dt=-T\Pi$, which follows from (\ref{87}).
Since $\Pi\ge0$, then $dF/dt\leq0$ and the free energy 
decreases monotonically in time towards its equilibrium
value. It is satisfying to realize that this inequality can
be regarded as the H theorem of Boltzmann. Indeed, if we define the
H function of Boltzmann by
\beq
H = \int \rho \ln \frac{\rho}{\rho_e} dx,
\eeq
and recalling that $\rho_e$ is proportional do the exponent
of $-\beta E$, we see that $H$ equals $-\beta F$, except for
an additive constant, a relation giving the result
\beq
\frac{dH}{dt} = - \beta \frac{dF}{dt} = k \Pi \geq 0,
\eeq
which is understood as the H theorem of Boltzmann.

\subsection{Microcanonical setting}

The second relevant case is the one in which $\rho_e$
vanish unless $E(x)=E_0$, which corresponds to the Gibbs 
microcanonical distribution. This condition is met if
the  left and right hand sides of the equation (\ref{90}) 
vanish, which give the conditions
\beq
D_i = \frac12 \sum_j \frac{\partial\Gamma_{ij}}{\partial x_j},
\label{78b}
\eeq
and 
\beq
\sum_j \Gamma_{ij} \frac{\partial E}{\partial x_j} = 0.
\label{78c}
\eeq
The covariances obeying this relation is obtained from the special
type of noise that we have analyzed above. 

Replacing result (\ref{78b}) into the expression (\ref{88a}),
we see that $L_i$ vanishes identically and so does the flux of
entropy, given by (\ref{70a}).

The heat flux also vanishes. To see this, it suffices to observe that
the covariances and dissipative forces, characterized by equations
(\ref{78b}) and (\ref{78c}), yields
\beq
J_i = -\frac12 \sum_j \Gamma_{ij} \frac{\partial \rho}{\partial x_j},
\label{94}
\eeq
which replaced in the expression (\ref{91b}) and making use of relation
(\ref{78c}) gives the vanishing of $\Phi_q$.  Thus not only the
flux of entropy is absent but also the heat flux, confirming that
the noise characterized by (\ref{78b}) and (\ref{78c}) describe
an isolated system. 

The insertion of the expression (\ref{94}) into the equation (\ref{24a})
gives the evolution in the form
\beq
\frac{\partial\rho}{\partial t}
= \{E,\rho\} + \frac12\sum_{ij}\frac{\partial}{\partial x_i}\Gamma_{ij}
\frac{\partial \rho}{\partial x_j},
\label{24b}
\eeq
and describes an isolated system as we have demonstrated.
In this sense it is similar to the Liouville equation 
\beq
\frac{\partial\rho}{\partial t} = \{E,\rho\},
\label{24c}
\eeq
that describes a isolated system. However, in the case of the
Liouville equation, the entropy is strictly constant in time,
and there is no entropy production. This is in contrast with
thermodynamic law of the increase of entropy in isolated systems,
but in agreement with the equation (\ref{24b}), which will generate
entropy. The variation of the entropy, which equals the rate of
entropy production $\Pi$, is given by
\beq
\frac{dS}{dt} = \frac{k}2 \sum_{ij} \int
\frac{\Gamma_{ij}}\rho\frac{\partial \rho}{\partial x_j}
\frac{\partial\rho}{\partial x_i} dx,
\label{93}
\eeq
which is clear nonnegative because $\Gamma_{ij}$ has nonnegative
eigenvalues, and we conclude that $dS/dt\geq0$.

\section{Mechanical system}

\subsection{General equations}

Here we apply the results obtained previously to a mechanical system
composed by a certain number of interacting particles with equal masses.
The positions of the particles are denoted by $x_i$, understood as even
variables, and the momenta of the particles by $p_i$, understood as odd
variables. The discrete time equations of motion are
\beq
x_i' = x_i + \frac{p_i}{m}\tau,
\label{101}
\eeq
\beq
p_i' = p_i + F_i\tau + D_i\tau + \xi_i \sqrt{\tau}, 
\label{102}
\eeq
where $F_i(x)$ is a conservative force that depends only on $x$,
that is,  $F_i=-dV/dx_i$, and $D_i$ is the dissipative force. 
The conservative force $F_i$ and $p_i/m$ hold the property (\ref{67a}),
as desired, and the dissipative force is assumed to hold the property
(\ref{67b}), which reads $D_i(x,-p)=-D_i(x,p)$. 

The equation (\ref{24a}) that gives the time evolution
of the probability density $\rho(x,p)$ reads
\beq
\frac{\partial \rho}{\partial t} = \{{\cal H},\rho\}
- \sum_i\frac{\partial J_i}{\partial p_i},
\label{17}
\eeq
where ${\cal H}$ is the energy function
\beq
{\cal H} = \sum_i \frac{p_i^2}{2m} + V(x).
\label{15}
\eeq
and we recall that $F_i=-\partial {\cal H}/\partial x_i$ and $p_i/m
=\partial {\cal H}/\partial p_i$.

We analyze initially the ordinary case in which the dissipative force
is proportional to the momentum, $D_i=-\gamma p_i$, and the covariances
are diagonal and do not depend on $x$ nor on $p$, and are given by
$\Gamma_{ii}=2m\gamma/\beta_i$. In this case the quantity $J_i$ is 
\beq
J_i = -\gamma \left(p_i \rho + \frac{m}{\beta_i}
\frac{\partial \rho}{\partial p_i}\right).
\label{17b}
\eeq
Replacing in equation (\ref{17}), the evolution equation reads
\beq
\frac{\partial \rho}{\partial t} = \{{\cal H},\rho\}
+ \gamma\sum_i\frac{\partial p_i \rho}{\partial p_i} 
+ \gamma m \sum_i \frac1{\beta_i} \frac{\partial^2\rho}{\partial p_i^2},
\label{92}
\eeq
which we recognize as the Fokker-Planck-Kramers equation for many
particles.

If $\beta_i=\beta$ is the same for all $i$, the noise-dissipation
relation is obeyed for the Gibbs probability density $\rho_e$
proportional to $e^{-\beta{\cal H}}$ and the equation (\ref{92})
describes a system in contact with a reservoir at a temperature
$T=1/k\beta$. For long times the system relax to the equilibrium state.
If $\beta_i$ are distinct, then for long times the system reaches
a nonequilibrium stationary state because $J_i$ cannot be zero for all $i$
and $\Pi\neq0$. In this case the equation can be understood as describing
a system in contact with several heat reservoirs at temperatures
$T_i=k\beta_i$. 

Another situation is the one in which $\rho_e$ vanishes unless
$H(x,p)=E_0$, which we have discussed above, and understood as describing
an isolated system. In equilibrium, it leads to the Gibbs microcanonical
distribution. In the present case where the equation of motion is given
by (\ref{101}) and (\ref{102}), the covariances are related only to the
momentum variable, so that the relation (\ref{78c}) gives
\beq
\sum_j \Gamma_{ij} p_j = 0.
\label{28c}
\eeq

The solution for $\Gamma_{ij}$ is
\beq
\Gamma_{ij} = -\lambda_{ij} p_i p_j \qquad\qquad i\neq j,
\label{28e}
\eeq
\beq
\Gamma_{ii} = \sum_{j{\neq i}} \lambda_{ij}p_j^2,
\label{28d}
\eeq
where $\lambda_{ij}=\lambda_{ji}\geq0$, which replaced into
(\ref{78b}) gives again the usual form of the dissipative force
\beq
D_i = - \gamma_i p_i, \qquad\qquad 
\gamma_i=\frac12\sum_{j(\neq i)} \lambda_{ij}.
\label{28b}
\eeq
The explicit form of $J_i$ is
\beq
J_i = \frac12 \sum_{j(\neq i)}\lambda_{ij}p_j
\left(p_i\frac{\partial\rho}{\partial p_j}
- p_j\frac{\partial\rho}{\partial p_i}\right).
\label{98}
\eeq

The flux of entropy $\Phi$ vanishes identically and the
time variation of entropy $dS/dt$ equals the rate of 
entropy $\Pi$. Using expression (\ref{93}),
we find
\beq
\frac{dS}{dt} = \frac{k}2  \sum_{i<j}\lambda_{ij}
\int \frac1\rho\left(p_j\frac{\partial \rho}{\partial p_i}
- p_i\frac{\partial \rho}{\partial p_j} \right)^2 dx,
\eeq
and we may conclude that $dS/dt\geq 0$.

\subsection{Weakly interacting particles}

As an example of a system that evolves with strictly constant energy,
we consider a system of weakly interacting particles in which case
the energy function can be taken as being just the kinetic energy,
\beq
{\cal H} = \sum_i \frac{p_i^2}{2m}.
\eeq
The evolution equation is
\beq
\frac{\partial \rho}{\partial t} = \{{\cal H},\rho\}
+ \frac12\sum_{ij}\frac{\partial}{\partial p_i}\Gamma_{ij}
\frac{\partial \rho}{\partial p_j},
\label{17a}
\eeq
where the covariances depend on $p_i$ according to
(\ref{28e}) and (\ref{28d}). 

As the energy function is strictly constant in time, the equilibrium
probability density is $\rho_e(p)$ is proportional to
$\delta({\cal H}(p)-E_0)$, as we have already seen. To solve equation
(\ref{17a}), we assume a probability distribution of the following form
$\rho(p) = g(p)\delta({\cal H}(p)-E_0) $, which we expect to be valid
near equilibrium, where
\beq
g = \frac1\zeta \exp\{- \frac12\sum_{i\neq j} b_{ij} p_i p_j \},
\eeq
and the quantities $b_{ij}$ are time dependent.
Replacing this form in the evolution equation we find the following
equation for $b_{ij}$, for $i\neq j$,
\beq
\frac{db_{ij}}{dt} = - \alpha_{ij}b_{ij},
\eeq
where
\beq
\alpha_{ij} = \lambda_{ij} + \frac12\sum_{k(\neq i)} \lambda_{ik} 
+ \frac12\sum_{k(\neq j)} \lambda_{jk}.
\eeq
The solution for $b_{ij}$ is
\beq
b_{ij} = c_{ij}e^{-\alpha_{ij} t},
\eeq
and we see that for long times the probability distribution decays
exponentially with time to the equilibrium distribution.

Let us determined the variation of entropy $dS/dt$, which for the
present case equals the rate of entropy production.
Using equation (\ref{93}), we find
\beq
\frac{dS}{dt} = \frac{k}{2}\sum_{i\neq j} b_{ij}^2
\{\lambda_{ij}[\la p_j^4 \ra - \la p_i^2\ra\la p_j^2 \ra] 
+ \sum_{k(\neq i)}\lambda_{ik} \la p_k^2\ra\la p_j^2 \ra\},
\eeq
where the averages are determined by using the equilibrium probability
distribution. We see that $dS/dt$ is positive and decays exponentially
to zero

I the probability density is only a function of the momenta, 
we see that the Poisson brackets in (\ref{17a}) vanishes
but that is not the case of last term on the right-hand side of 
(\ref{17a}). The vanishing of the Poisson brackets means that the
Liouville equation gives $\rho$ constant in time and thus do not
relax to the equilibrium solution, if it was out of equilibrium at
the beginning. This is in contrast with the solution
of equation (\ref{17a}) which predicts a relaxation to equilibrium 
and a nonzero production of entropy, and $dS/dt>0$ out of equilibrium.

\section{Conclusion}

We have developed an approach to stochastic thermodynamics of
systems with continuous space of states. The results were obtained
by continuous time limit of a discrete time formulation,
which includes the evolution equation and the rate of entropy
production. We have emphasized the role of the time reversal and 
of the dissipation probability
current in the properties related to irreversible processes.
When this part of the probability current vanishes, the rate
of entropy production vanishes, and the equilibrium sets in.
The rate of entropy production was shown to be a bilinear form
in the components of the dissipation probability current and
is positive definite.

We have also analyzed a type of noise that makes the energy function
to be strictly constant along a stochastic trajectory and thus
describing an isolated system. The increase in entropy is
entirely due to the generation of entropy inside the system.
This theoretical result is in agreement with thermodynamics 
in the sense that the entropy of an isolated system, in general,
increases. This result contrasts with the prediction given by 
the Liouville equation that the entropy is constant in time,
and there is not generation of entropy, although 
this equation describes an isolated system as the 
energy is strictly constant in time.


\end{document}